\documentclass[aps,prl,reprint,groupedaddress,twocolumn]{revtex4-1}
\usepackage[pdftex]{graphicx}
\usepackage{epstopdf}
\usepackage{bm}
\usepackage{amssymb}
\usepackage{psfrag}
\usepackage{color}
\usepackage{amsbsy}      
\usepackage{amsmath,amsthm}
\usepackage{mathtools}
\usepackage{comment}
\graphicspath{{./figures/}}
\usepackage{newfloat}
\DeclareFloatingEnvironment[name={Fig. D}]{suppfigure}

\usepackage{natbib} 
\newcommand{\bsigma}{{\boldsymbol\sigma}}

\def\d{{\bf d}}
\def\dd{\mbox{d}}

\def\f{{\bf f}}

\def\r{{\bf r}}

\def\u{{\bf u}}

\newcommand{\Conv}{\mathop{\scalebox{1.5}{\raisebox{-0.2ex}{$\ast$}}}}%
\usepackage[inline,final]{changes}
\usepackage{comment}

\newcommand{\remove}{\deleted}
\newcommand{\add}{\added}

\begin{document}

\title{High-resolution reconstruction of cellular traction-force
  distributions: the role of physically motivated constraints and
  compressive regularization}


\author{Joshua C. Chang$^{1}$, Yanli Liu$^{2}$, and Tom Chou$^{2,3}$}
\affiliation{$^{1}$Epidemiology and Biostatistics Section, Rehabilitation
  Medicine, Clinical Center, The National Institutes of Health,
  Bethesda MD, 20892} \affiliation{Dept. of Mathematics, UCLA, Los Angeles, CA 90095-1555}

\maketitle 


\section{Abstract}

We develop a method to reconstruct, from measured displacements of an
underlying elastic substrate, the spatially dependent forces that
cells or tissues impart on it.  Given newly available high-resolution
images of substrate displacements, it is desirable to be able to
reconstruct small scale, compactly supported focal adhesions which are
often localized and exist only within the footprint of a cell.  In
addition to the standard quadratic data mismatch terms that define
least-squares fitting, we motivate a regularization term in the
objective function that penalizes vectorial invariants of the
reconstructed surface stress while preserving boundaries.  We solve
this inverse problem by providing a numerical method for setting up a
discretized inverse problem that is solvable by standard convex
optimization techniques. By minimizing the objective function subject
to a number of important physically motivated constraints, we are able
to efficiently reconstruct stress fields with localized structure from
simulated and experimental substrate displacements.  Our method
incorporates the exact solution for the stress tensor accurate to
first-order finite-differences and motivates the use of distance-based
cut-offs for data inclusion and problem sparsification.

\section{Introduction}

The adhesion of cells and tissues to their environment has profound
consequences on processes such as cell polarization \cite{MASHA2011},
division, differentiation \cite{DIFFERENTIATION0}, tissue morphology
during development \cite{DEVELOPMENT0}, wound healing
\cite{WOUND0,WOUND2,WOUND1}, and cancer metastasis
\cite{CANCER0}. Hence, quantifying how cells attach to impart force
on the surrounding material is an important technical challenge in
cell biology.

\begin{figure}[t]
\begin{center}
\includegraphics[width=3.3in]{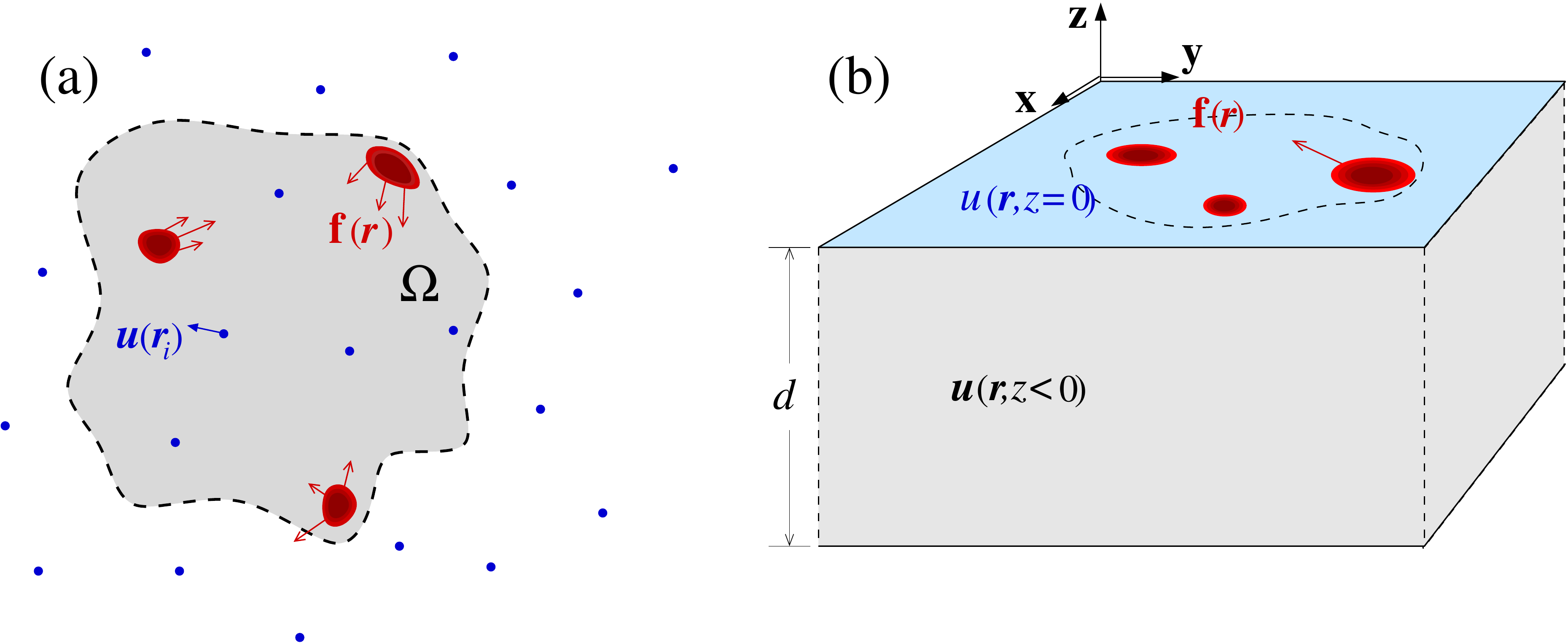}
\caption{\textbf{A schematic of an isolated cell.} (a) The boundary of the cell
  footprint is denoted by the dashed curve, the support of the stress
  field is represented by the red regions that impart a stress ${\bf
    f}(x,y)$ on the surface. Displacements ${\bf u}(\r_{i})$ of the
  elastic medium are measured at position $\r_{i} =
  x_{i}\hat{x}+y_{i}\hat{y}+z_{i}\hat{z}$ (blue dots) that can be
  inside or outside the cell footprint, on the surface ($z_{i}=0$), or
  below the surface ($z_{i}<0$). (b). A perspective view of the
  elastic substrate and cellular footprint.}
\label{FIG1}
\end{center}
\end{figure}

Cell motility and cellular response to signals have hitherto typically
been studied in two-dimensional geometries in which cells are placed
on a flat elastic substrate.  Dynamic adhesion between the cells and
the substrate are realized through dynamically reorganizing focal
adhesions, often mediated through cellular structures such as
lamellipodia and filopodia \cite{MBOC}. Focal adhesions are typically
spatially localized, as shown in Fig.~\ref{FIG1}. Similarly, on larger
length scales, a collection of cells can give rise to localized stress
distributions. For example, the leading edge of a cell layer produces
the pulling force that leads to migration in wound healing assays
\cite{WOUND0,WOUND1,WOUND2}.

Dynamically varying force-generating structures are often small and
difficult to image. Mechanics-based methods for inferring their
positions and magnitudes, such as using deformation of pillar
structures \cite{PILLAR} or textured substrates \cite{PATTERN}, have
been developed.  These methods require the cell to attach to a
non-flat interface.  The simplest method compatible with a flat
interface relies on measuring the displacement of fiduciary markers,
such as gold nanoparticles, embedded in the elastic
substrate~\cite{WANG2007}. The measured displacements are an indirect
probe of the force-generating structures, \textit{e.g.}, focal
adhesions.  Any inversion method should be able to not only
reconstruct the positions and magnitudes of the stress field, but
should ideally be able to capture potentially sharp boundaries of the
stress-generating structures. However, fiduciary markers embedded in
the 3D substrate are typically too sparse to reveal a displacement
field with sufficient resolution to infer small cellular focal
adhesion structures. To image such sub-cellular stress structures,
high resolution reconstructions are required \cite{USCHWARZ,YORK}.
Experimentally, new high-resolution imaging methods have been
developed using methods to image higher densities of fiduciary markers
\cite{YORK} or fluorescent grid patterning of the substrate
\cite{POPESCU}. A surface grid pattern of fluorescent adhesion
proteins allows surface deformation to be directly measured using
conventional microscopes.

In light of such higher spatial resolution techniques, we develop an
improved method for elastic stress source recovery using ideas
developed for image segmentation \cite{OSHER}.  This class of methods
relies on optimization that uses ``compressive'' $L^{1}$
regularization terms in the objective function that favor solutions
that are compactly supported \cite{CHAN,DANUSER}.  This type of
regularization term is not derived from any fundamental physical law,
but represents prior knowledge that the function to be recovered is
sparse in content\deleted{ except near edges}. In addition, the overall
objective function will be constructed to obey physical constraints
and symmetries.

In the next section, we review the standard linear equations of
elasticity that describe the forward problem of computing the
displacement field as a function of an arbitrary surface stress
distribution. This model is then used to construct the data mismatch
term in the objective function. We then motivate regularization and
constraint terms in the full objective function. Finally, we
demonstrate our method using both simulated and experimental data. Our
method provides good reconstruction of localized structures that
exhibit desirable qualities such \added{compressive recovery of
compact features} as \added{well as } the suppression of Gibbs ringing
phenomenon at the boundaries of the stress structures.

\section{Methods}

\subsection{Forward problem}

We first derive the linear elastic Green's function associated with a
point force applied to the surface of a semi-infinite half-space, as
shown in Fig.~\ref{FIG1}(b). We assume that the elastic medium is
infinite in both depth ($d\to \infty$) and lateral extent. The Green's
function tensor defined in the domain ${\cal D}=\left\{(x,y,z)|x,y\in
R,z\leq0\right\}$ is given by
\begin{equation}
{\bf G} = \left[ \begin{matrix} G_{xx}(x,y,z) & G_{xy}(x,y,z) & G_{xz}(x,y,z) \\
	G_{yx}(x,y,z) & G_{xy}(x,y,z) & G_{yz}(x,y,z) \\
	G_{zx}(x,y,z) & G_{zy}(x,y,z) & G_{zz}(x,y,z) 
 \end{matrix} \right]
\end{equation}
where the components are explicitly given in 
Appendix A of the Supplementary Material. For example, 

\begin{equation}
G_{sz, zs}(x,y,z) =
\frac{1+\nu}{2\pi E}\left(\frac{sz}{R_{\perp}^{3}}\pm\frac{(1-2\nu)s}{R_{\perp}
(R_{\perp}-z)}\right).
\end{equation}
where $s\equiv x,y$. The equation with $\pm$ corresponds to $G_{sz}$
and $G_{zs}$, respectively, and $R_{\perp} \equiv \sqrt{x^{2}
  +y^{2}}$. The Young's modulus and Poisson ratio of the elastic
substrate are denoted by $E$ and $\nu$, respectively.  For Matrigel,
$E\approx4\pm3\times10^2\textrm{ Pa}$ and
$\nu\approx0.5$~\cite{SOOFIA2009}. Throughout the rest of this
manuscript, we will express stress in units of $E$.  The displacement
of a material point at $(x,y,z\leq 0)$ in the medium due to a stress
distribution ${\bf f}$ is simply the convolution $\u(\r) \equiv [u_x
  \ u_y\ u_z]^\intercal = {\bf G}\Conv\f$.
%

%
%
%
%
%

For our specific problem, we shall restrict the forces to surface
stresses $\f$ that act on the plane perpendicular to the
$\hat{z}$ axis. We define the in-plane stress distribution, at depth $z$, as
$\f(x,y) = f_{x}(x,y)\hat{x} + f_{y}(x,y)\hat{y}$. 
The resulting surface-level displacement fields become

\begin{equation}
u_{s}(x,y,z) = \sum_{k=x,y,z}\int_\Omega \dd x'\dd y'G_{sk}(x-x',y-y',z)f_{k}(x',y').
\label{eq:UMODEL1s}
\end{equation}
%
Note that tangential stresses can induce
displacements in the direction normal to the surface.
For cells on flat surface, we assume that $f_{z}=0$.

\subsection{Inverse problem}

Next, we develop an objective function for which the minimizing
solution provides a good approximation to the underlying stress field,
while preserving discontinuities.  The first component is simply a
quadratic data mismatch term defined by the sum over the displacements
measured at the $N$ measurement positions at $\r_{i}$:

\begin{equation}
\Phi_{\rm data}[\f] = \sum_{i}^{N}\vert \u^{\rm
  data}(\r_{i})- \u(\r_i\vert \f)\vert^{2}.
\label{PHIDATA}
\end{equation}
Since $\u^{\rm data}(\r_{i})$ is given, and $\u(\r_{i}\vert \f)$ is
given by the linear model of Eq.~\ref{eq:UMODEL1s}, this contribution
to the objective function is a functional over the surface force
$\f(x,y)$.  For simplicity, we will assume that the data points are
measured only at the interface $z=0$ over an uniform grid with
coordinates given $\{ (x_j,y_k) : j\in\{1,2,\ldots,J\},
k\in\{1,2,\ldots,K\}\}.$

In Eq.~\ref{eq:UMODEL1s}, we have restricted the domain of integration
to lie within the cell footprint $\Omega$, reiterating that $\f(x,y)$
has compact support. As a consequence of compact support, for a fixed
discretized approximation of $\f(x,y)$, the displacements can be
obtained exactly by solving an equivalent system of linear equations
of finite dimension. We explicitly define this system of linear
equations given a piecewise-affine approximation of the stress
field. Let us consider the first-order approximation of $f_{x}(x,y)$
and $f_{y}(x,y)$ using central finite differences, for $x\in[x_j -
  \delta x/2, x_j+\delta x/2) \cap y\in[y_j-\delta y/2, y_j + \delta y
    /2)$,
\begin{align}
\lefteqn{f_{x}(x,y) =f_{x}(x_i , y_j)  }\nonumber \\
& \qquad+ (x-x_i)\frac{f_{x}(x_{i+1},y_j) -f_{x}(x_{i-1},y_j) }{2\delta x}  \nonumber\\
&\qquad + (y-y_j)\frac{f_{x}(x_i,y_{j+1}) - f_{x}(x_i,y_{j-1}) }{2\delta y} \nonumber\\
&\qquad + \mathcal{O}(\delta x)^2 + \mathcal{O}(\delta y)^2,\label{eq:sigma_affine}
\end{align}
where $i,j$ denotes a tuple of grid coordinates. In effect, we are
performing sub-pixel interpolation \deleted{of the stress} where the stress is
fully-determined by its values at the grid vertices.

Upon using Eq.~\ref{eq:sigma_affine}, we can rewrite
Eq.~\ref{eq:UMODEL1s} by decomposing the integral into a sum of
integrals over grid cells.  After further regrouping terms, we find a
linear system of equations for $u_{s}(x,y)$ across all grid points. For example, 

\begin{equation}
u_{x}(x_{n},y_{m}) = \Gamma_{xx}^{nmjk}f_{x}(x_j,y_k) + \Gamma_{xy}^{nmjk}f_{y}(x_j,y_k),
\label{eq:linearsystem1}
\end{equation}
where summation notation for each index tuple $(j,k)$ has been
implicitly assumed, and the tensors $\Gamma_{xx}^{nmjk}$, $\Gamma_{xy}^{nmjk}$, and the
analogous formulae for $u_{y}(x_{n},y_{m})$ are given in Appendix B of
the Supplementary Material.

From an equation-counting perspective, the system of equations is
exactly determined \replaced{if and only if}{given that} one has at least as many measurement
points as grid cells in the resolution that one wishes to reconstruct
the stress field, provided that one is able to measure displacements
in both principle directions. Even if one is able to measure \remove{both}
displacements \add{in both directions}, \add{the measurements may be 
imprecise and noisy rendering the inversion
of Eq.}~\ref{eq:linearsystem1} \add{highly ill-conditioned.} 
\remove{the problem may still be difficult since the inversion
of Eq.}~\ref{eq:linearsystem1} \remove{maybe highly ill-conditioned and the
measurements are taken in the presence of noise at a finite
precision.}
To resolve these issues, we introduce a number of
physically consistent constraints and regularization terms relevant to
this system.

\subsection{Physical constraints and regularization}

The remaining components of the objective function should contain
information about the known physical constraints as well as
regularization terms that better condition the overall optimization
problem. Various regularization terms have been motivated, but they
can also be associated with prior knowledge on the solution
\cite{PATHINTEGRAL}. 

First, we consider explicit physical constraints. Since we are
assuming inertial effects are negligible, we require that the net
force vanish, or that
\begin{equation}
\int_\Omega f_{x}(x,y)\dd x \dd y= \int_\Omega f_{y}(x,y)\dd x \dd y = 0.
\label{NOFORCE}
\end{equation}
Likewise, we require that there is no net torque, or that
\begin{equation}
\int_\Omega x f_{y}(x,y) \dd x \dd y  = \int_\Omega y f_{x}(x,y) \dd x \dd y.
\label{NOTORQUE}
\end{equation}
Similar no-force and no-torque constraints have been previously
applied to the traction force inference problem in the Fourier domain
\cite{BUTLER2002} for which sparse solutions are difficult to resolve.

Another physical constraint is the requirement that surface stress at
locations outside of the cell footprint vanish.  In regions where
there is no contact between the cell and the substrate, no mechanism
can impart stress.  Thus, the stress field is compactly supported
within the cell footprint.  The stress field may be further localized
within cellular focal adhesions inside the cell footprint. Confining
the stress to within an arbitrary cell footprint requires a complex
iterative method \cite{BUTLER2002}.

To better condition the inference of $\f(x,y)$, we regularize this
problem by forcing the reconstruction to obey some physically relevant
characteristics of the surface stress. In many other types of inverse
problems, for example, in the inference of the potential of mean force
of a molecular bond, a constraint on differentiability is typically
imposed on the function to be inferred \cite{PATHDFS}. A typical
constraint of this nature may be a quadratic penalty on the gradients
of the function to be inferred.  However, such $L^{2}$ functional
regularization often leads to \replaced{over-smoothing of extreme values
and failure to recover compactness}{ringing and inaccurate functional
reconstruction, especially when the underlying function is highly
localized or compactly supported}. Thus, $L^{1}$ regularization on the
function, or on its variations, have been developed to allow for
more ``compressed'' reconstructions. These approaches
are suitable for problems such as segmentation of images where
boundaries are are sharp \cite{CHAN}.  In these problems, the data is
sparse in the sense that the boundaries in an image contain most of
the information.  Likewise, in the stress recovery problem at hand,
the data consisting of displacements at a finite number of measurement
positions may be considered sparse.

To this end one may employ variants of penalties often used in
image processing applications, where one penalizes the $L^1$
norm of the vector field or its variations using regularization terms 
$\Phi_{\rm reg}$ of the form
\begin{equation}
\Phi_{L^{1}} = \int_\Omega \left(\vert f_{x}(x,y)\vert  +\vert f_{y}(x,y)\vert\right) \d\r,
\label{L10}
\end{equation} 

\begin{equation}
 \Phi_{{\rm TV}_1} = \int_\Omega \left( \vert\nabla f_{x}(x,y) \vert +
 \vert\nabla f_{y}(x,y) \vert\right)\d\r
 \label{eq:PhiTV1}
\end{equation}
and
\begin{equation}
\Phi_{{\rm TV}_2} = \int_\Omega \left(\vert\partial_x f_{x}\vert 
+ \vert\partial_y f_{x}\vert + \vert\partial_x f_{y}\vert
+ \vert\partial_y f_{y}\vert\right)\d\r,
 \label{eq:PhiTV2}
\end{equation}
representing an $L^{1}$ regularization of the surface stress and two
forms of its total variation, respectively.  Since these
regularization terms are not based on any fundamental physical law,
there is some freedom \remove{of} in choosing their form.  However, we do not
want the choice of parameterization for the data grid to affect the
reconstruction results. Hence, the regularization terms should not
induce any additional anisotropy over that of the measured displacement field.
Thus, appropriate regularizations should be invariant under coordinate
rotation. Coordinate-invariant regularizers can be constructed
from the magnitude of the force vector at the surface
\begin{equation}
\vert\f(x,y) \vert  = \sqrt{f_{x}^{2} + f_{y}^{2}}.
\end{equation}
Any regularization penalty imposed on the reconstruction problem must
be a functional of this quantity in order to maintain rotational
invariance relative to the choice of how the displacements are
sampled.  In this manuscript, we follow the approach taken by Han
\textit{et al.} \cite{DANUSER} and focus on the isotropic $L^1$ norm
\begin{equation}
\tilde{\Phi}_{L^{1}} = \int_\Omega \vert\f(x,y) \vert \dd x\dd y.
\label{eq:mag}
\end{equation}
%
Other regularizations are possible; for example, one may also use the isotropic $L^2$ norm
\begin{equation}
\tilde{\Phi}_{\rm L^2} =  
 \int_\Omega \vert\f(x,y) \vert^2 \dd x\dd y,
\label{eq:L2}
\end{equation}
as was used by Plotnikov \textit{et al.} \cite{WATERMAN}.

Employing any of the above expressions as the regularization norm
$\Phi_{\rm reg}$, we define the penalized optimization problem

\begin{equation}
\hat{\f} \big\vert \lambda = \arg\min_{\f} \left\{ \Phi_{\textrm{data}}[\f] + 
\lambda\Phi_{\textrm{reg}}[\f] \right\},
\label{eq:objective}
\end{equation}
subject to the no-force, no-torque, and footprint constraints
(Eqs.~\ref{NOFORCE} and \ref{NOTORQUE}) on $\f$ mentioned above.
In Eq.~\ref{eq:objective}, $\lambda>0$ is a tunable parameter. This
problem is in a standard form that is directly solvable using a
variety of optimization routines.  In our implementation, we use a
second-order quadratic cone solver~\cite{cvxpy}.

To reduce the size of the system of equations described in
Eqs.~\ref{eq:linearsystem1}, we note that the Green's function falls
off at a rate of $|\r|^{-1}$. However, when combined with the
zero-force constraint, the relationship between the
displacements and the support of the stress field falls off at the
much quicker rate of $|\r|^{-2}$ (see Appendix C in Supplementary Material). Formally, if 
%
%
$\Omega = {\rm sup}(\bsigma)\subset \mathbb{R}^2$ is compact, and
$\int \bsigma(\r)\d\r = \mathbf{0}$, then as $\mathbf{r}\to\infty$,
$u_{x,y}(\r) = \mathcal{O}\left( |\r|^{-2} \right)$.  The decay of the
influence of stress on the system provides justification for setting
distance-based cut-offs of the linear system. The effect of the
cut-off is to limit the left-hand side of Eq.~\ref{eq:linearsystem1}
to locations only within some maximal distance $R_{\perp}$ from the
outline of the cell.


\section{Results}

We implemented our regularized inversion method in Python version 3.5,
where optimization is performed using the \texttt{cvxpy} package with
the \texttt{ecos} solver.  Our implementation is available at
\url{https://github.com/joshchang/tractionforce}.  For all reconstructions,
we assumed that $\nu=0.5$ and reported all results in normalized
units of the Young's modulus.

\begin{figure*}
\begin{center}
\includegraphics[width=5.6in]{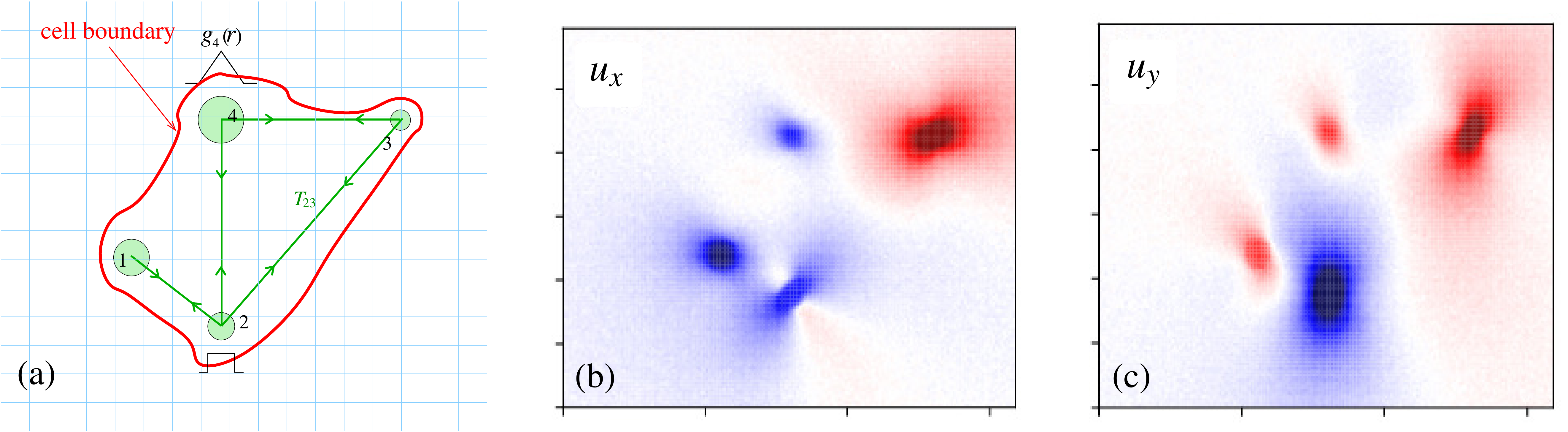}
\caption{\textbf{Test stress pattern and surface displacements.} (a)
  Four focal adhesions attached by filaments indicated by the green
  lines. The red border represents the extent of the cell footprint
  and can be determined experimentally as part of the imaging.
  Mathematically, the cell boundary forms the basis for a constraint
  on the stress distribution and we explore the dependence of the
  quality of reconstruction on the footprint constraint. The faint
  blue grid represents the regular points at which displacements might
  be measured. (b) and (c) show, schematically, the corresponding surface
  displacements $u_{x}(x,y)$ and $u_{y}(x,y)$,
  respectively.\label{TEST}}
\end{center}
\end{figure*}

\subsection{Simulated data}

First, we tested our method on simulated data derived from a force-
and torque-free test stress field shown in Fig.~\ref{TEST}. The test
pattern consists of four separated circular stress pads, or focal
adhesions, with radii $r_{1} = 1/5$, $r_{2} = 1/6$, $r_{3} = 1/8$, and
$r_{4} = 1/4$, and centers at positions $(x_{1},y_{1}) = (-1,-1/2)$,
$(x_{2},y_{2}) = (0,-1)$, $(x_{3},y_{3}) = (2,1)$, and $(x_{4},y_{4})
= (0,1)$.  The pads 2,3, and 4 are connected in a triangle as shown,
while pad 1 is connected only to pad 2.  The tensions along these
connections give rise to surface stresses imparted by the pads onto
the substrate.  We will assume that the stress fields in pads 1,2, and
3 are uniformly distributed within the circular disks. For pad 4, we
assume that the filaments connected to it \add{are} distributed according to a
cone-like density function. Thus, the stress field within pad 4
linearly decrease along the radial direction. The stresses $\f^{(i)}$
under each patch $i$ are decomposed into contributions arising from
the total tension $T_{ij}$ connecting them with pad $j$ and can be
expressed in the form

\begin{align}
\f^{(1)} & = a_{12}\left(\hat{x} -{\hat{y}\over 2}\right) \label{SIGMA1}\\
\f^{(2)} & = {G_{4}\over A_{2}}\hat{y} - a_{12}{A_{1}\over A_{2}}\left(\hat{x} 
-{\hat{y}\over 2}\right) + a_{23}(\hat{x}+\hat{y}) \label{SIGMA2} \\
\f^{(3)} & = -\hat{x}{G_{4}\over A_{3}} - a_{23}{A_{2}\over A_{3}}(\hat{x}+\hat{y}) 
\label{SIGMA3}\\
\f^{(4)} & = g_{4}\left(1- {r \over r_{4}}\right)(\hat{x}-\hat{y})\label{SIGMA4}
\end{align}
where $a_{12}, a_{23}, g_{4} >0$ are constant amplitudes, 
$A_{i} = \pi r_{i}^{2}$ are the pad areas, and 

\begin{equation}
G_{4} = 2\pi g_{4}\int_{0}^{r_{4}}\left(1-{r\over r_{4}}\right)r\dd r = {g_{4}\pi r_{4}^{2}\over 3}
\end{equation}
is the total force on pad 4 in each direction. Note that both test
stress fields are constructed to be force- and torque-free.

\begin{figure*}
\includegraphics[width=\linewidth]{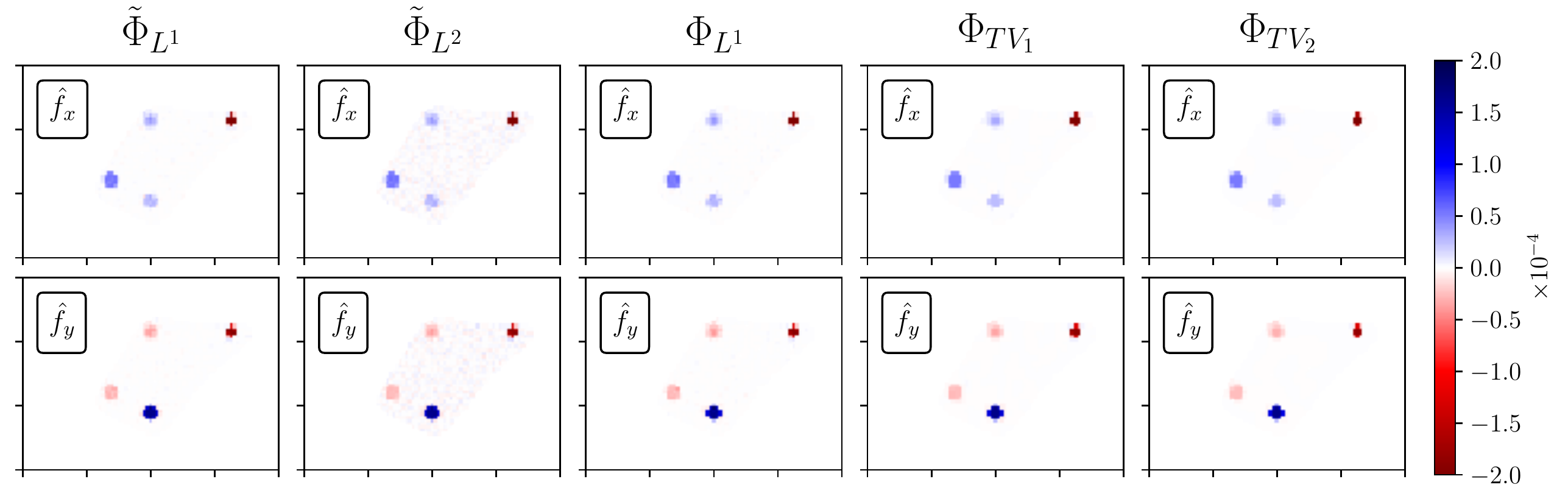}
\caption{\textbf{Comparing different regularizers.} Reconstruction of
  the test patterns using all constraints and the different forms of
  $\Phi_{\rm reg}$: $\Phi_{L^{1}},\Phi_{{\rm TV}_1},\Phi_{{\rm
      TV}_2}$,$\tilde{\Phi}_{L^{1}}$ and $\tilde{\Phi}_{\rm
    {L}_{2}}$. Regularization parameters chosen according to
  Fig.~D1 in Appendix D of the Supplementary Material.}
\label{COMPARE}
\end{figure*}

In our examples, we used $a_{12} = 10^{-4}$, $a_{23} =
2\times10^{-4}$, and $g_4 = 9\times10^{-5}$.  We generated
displacement fields by solving the forward problem of (\textit{i.e.},
Eq.~\ref{eq:UMODEL1s} for surface values $\u(x,y,z=0)$) and then
corrupted the displacements with Gaussian white noise with standard
deviation $10^{-5}$. From these noisy displacements, we reconstructed
$f(\r)$.

\begin{figure*}
\includegraphics[width=4.4in]{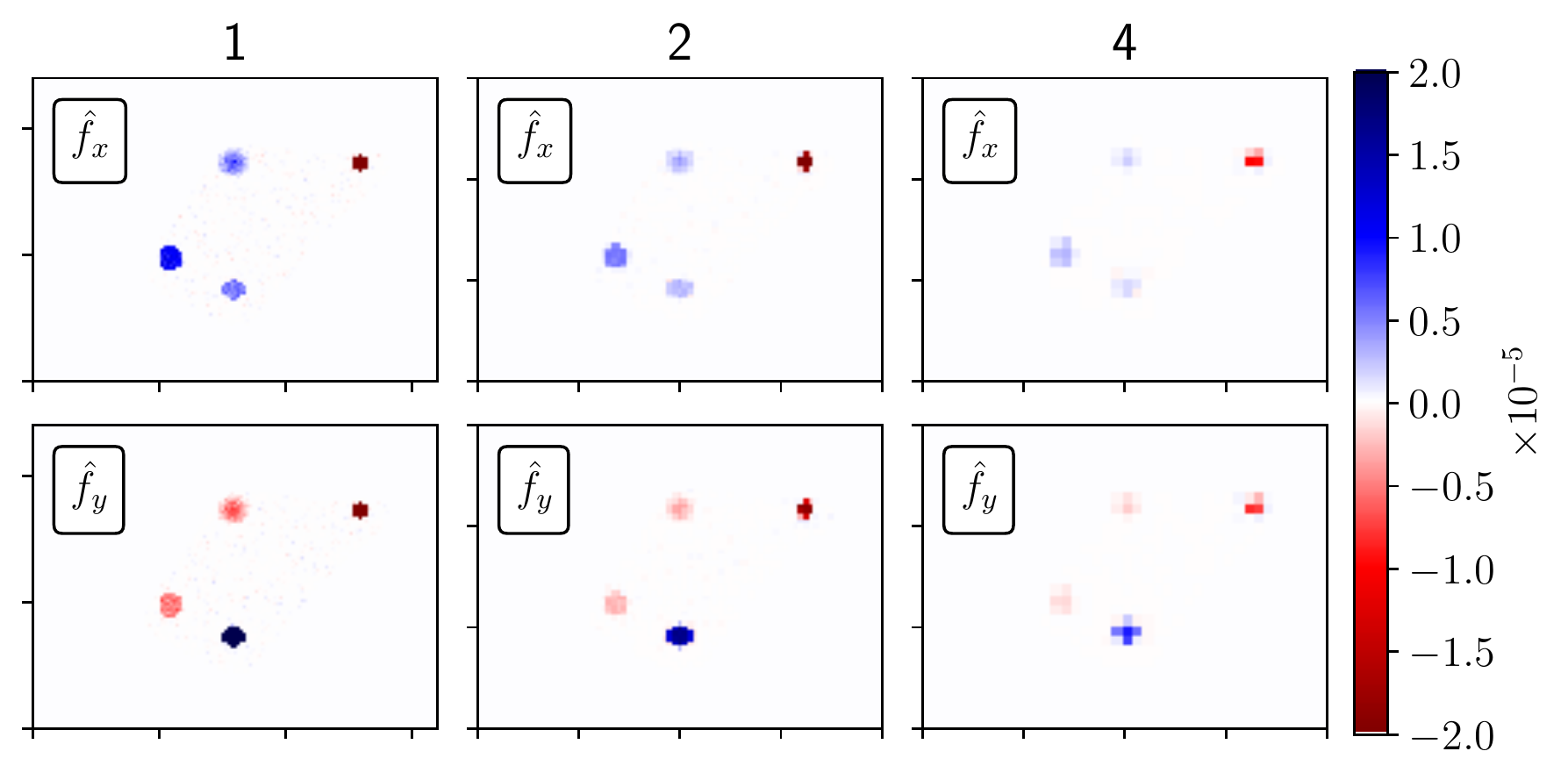}
\caption{\textbf{Grid coarsening.} Using every $n\in\{1,2,4\}$ lattice
  points of observations. The reconstruction is also performed at the
  same resolution.  In general, the optimization is stable and the
  qualitative features of the reconstructed $\f(x,y)$ are robust to
  modest data coarsening.}
\label{GRID}
\end{figure*}

Fig.~\ref{COMPARE} compares the reconstruction achieved from using
the different forms of $\Phi_{\rm reg}$. In all of these
reconstructions, we have imposed that the surface stress is both
force-free and torque-free, and also that the support of the surface
stress is within given the boundary defined in Fig.~\ref{TEST}.  The
adjustable parameter $\lambda$ was chosen in each instance by
examining the balance between data mismatch and regularity using
trade-off curves shown in Fig.~D1 in Appendix D, and taking
the value for $\lambda$ that yields a point farthest away from the
line segment joining the ends of the plot. The chosen value of
$\lambda$ corresponds to a balance between regularity and data
fidelity. The solution corresponding to each particular value of
$\lambda$ is shown in Fig.~\ref{COMPARE}.  Each \remove{row}\add{column} in
Fig.~\ref{COMPARE} corresponds to the use of a different
regularization penalty. The parameter $\lambda$ can also be extracted
using a Bayesian framework in which posterior probability is maximized
\cite{WATERMAN}.

Fig.~\ref{COMPARE} indicates that all forms of regularization yield
reasonable reconstructions of the four pads, albeit at different
levels of scarcity in reconstruction of the surrounding regions. The
reconstruction using the isotropic $L^1$ penalty is seen to be more
sparse than that of the isotropic $L^2$ penalty, whereas the other
penalties all yielded comparable reconstructions. In the remaining
analysis of the 4-pad test pattern we concentrate on using
$\tilde{\Phi}_{L^{1}}$.

\begin{figure}
\includegraphics[width=2.8in]{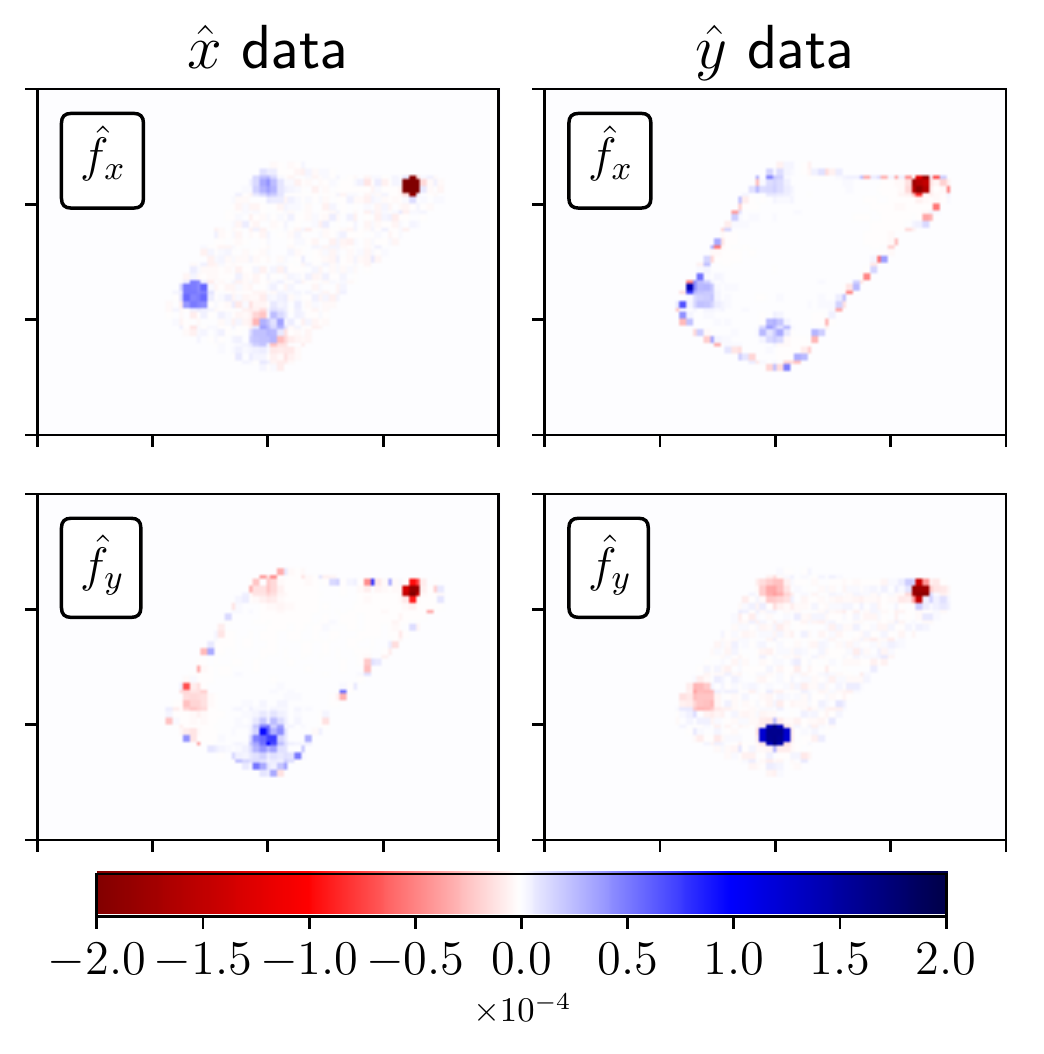}
\caption{\textbf{Unidirectional displacement measurements.} We explore
  surface stress reconstruction from displacements only in one
  direction. Reconstructions of both components of the four-pad
  surface stress shown under measures along only $\hat{x}$ or only
  $\hat{y},$ using the $\tilde{\Phi}_{L^1}$ norm.}
\label{XYONLY}
\end{figure}

In many inverse problems, computational complexity is a technical
issue either due to memory constraints or computational time. A tactic
for reducing computational complexity is to coarsen the reconstruction
problem so that one reduces the rank of the linear system to solve.
Fig.~\ref{GRID} shows reconstructions of the 4-pad test patterns
using $\tilde{\Phi}_{L^{1}}$ as a function of the coarseness of the
displacement data. We coarsened the data by taking only every
$n\in\{1,2,4\}$ lattice points in each dimension, noting that doing so
reduces the rank of the problem by a factor of $n^2$.  The general
features of the stress patterns are preserved under coarsening but
sufficient density of data points are needed to resolve fine scale
variations in the stress field.

\begin{figure*}
\includegraphics[width=4.4in]{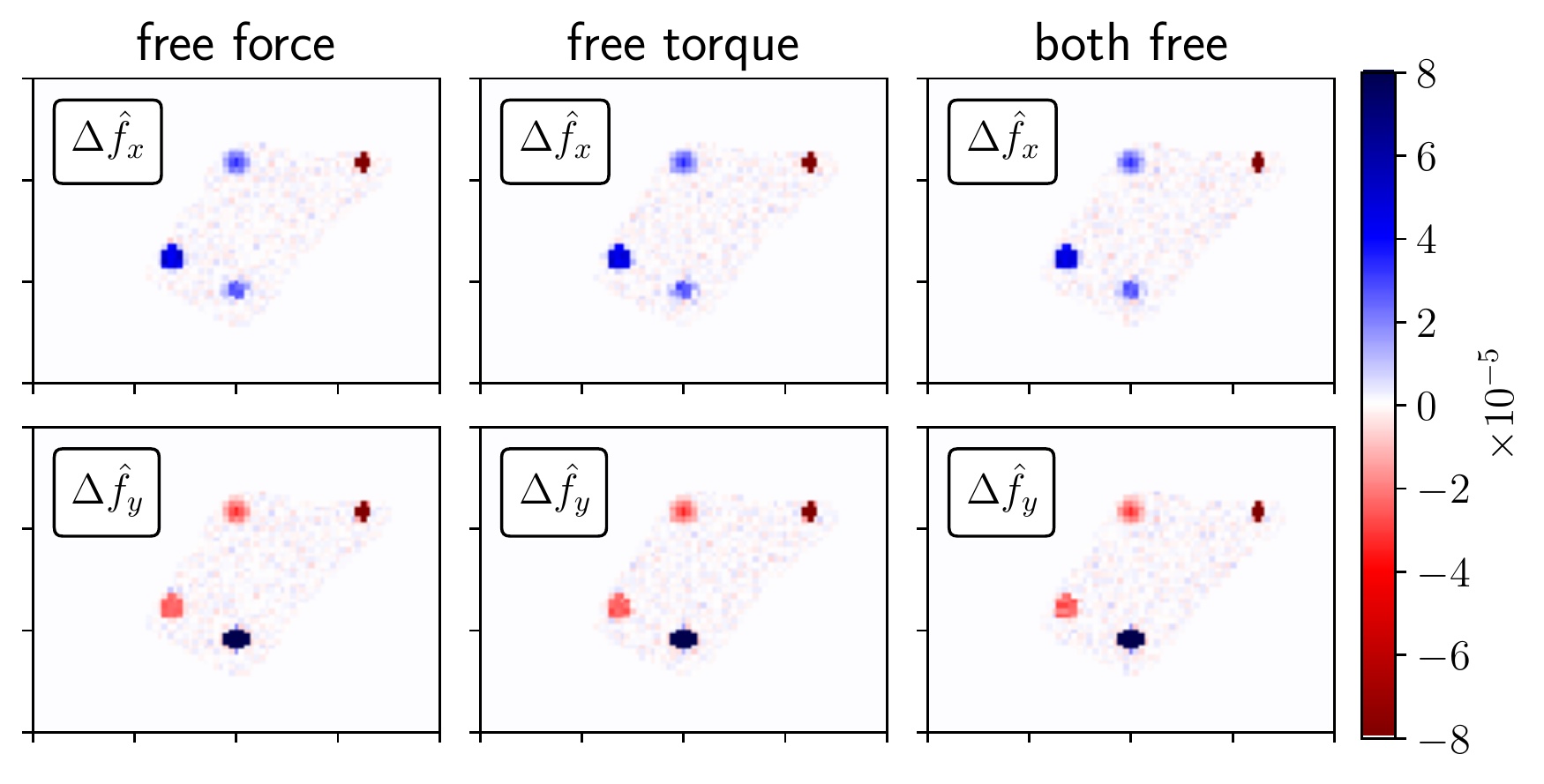}
\caption{\textbf{Constraints are unsatisfied unless enforced.} Plotted
  are best reconstructions under $\tilde{\Phi}_{L^{1}}$ penalty and
  the difference between these reconstructions and the corresponding
  fully constrained reconstruction in Fig.~\ref{COMPARE}. }
\label{CONSTRAINTS}
\end{figure*}

\begin{figure}[h!]
\includegraphics[width=2.9in]{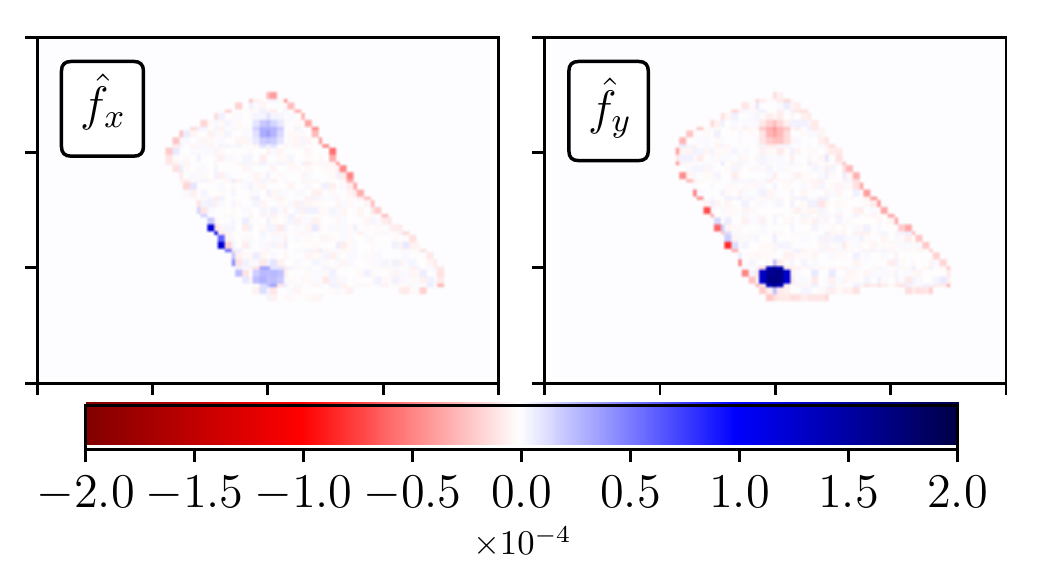}
\caption{\textbf{Footprint misspecification.} Reconstructions
  performed with the footprint drawn in an incorrect orientation.  The
  erroneous footprint constraint misses two of the stress pads,
  resulting in a erroneous forces distributed along the edge of the
  footprint. \add{If the misspecified footprint misses actual stress pads, 
spurious stresses will be generated at the misspecified boundaries in
order to satisfy the no-force and no-torque constraint.}}
\label{FOOTPRINT}
\end{figure}


In some applications, displacements are only measurable along a single
axis. Reconstruction of the surface forces using this type of data is
related to the problem of missing data or information loss.  We tested
reconstruction under these circumstances by assuming that
displacements only in either the $\hat{x}$ or the $\hat{y}$ directions
are measured. In Fig.~\ref{XYONLY} we show reconstructions using only
$u_{x}$ or $u_{y}$ data. In both cases, the reconstruction of the
force in the direction perpendicular to the measured displacements is
most affected by information loss. This effect is seen in the
\remove{presence of fictitious} \add{spurious} boundary forces
\add{generated at the cell boundary} in both cases. \remove{as the
  cell boundary is visible in these reconstructions.}
\remove{Yet}\add{Nonetheless}, the pads are clearly visible in both
sets of reconstructions, along both axes.


In all reconstructions so far, we have enforced the physical no-force
and no-torque constraints. Fig.~\ref{CONSTRAINTS} explores the
influence of these constraints by providing the differences in the
reconstructions between the physically constrained problem and
unconstrained problems. In the first \remove{row}\add{column}, the
force constraint is not enforced. In the second
\remove{row}\add{column}, the torque constraint is not enforced. In
the last \remove{row}\add{column}, neither force nor torque
constraints are enforced.  In all cases, we see that the constraints
are active; removing the constraints affect the quantitative results
of the reconstruction.  In all cases, constraints are not satisfied
automatically. For example, without enforcing the force constraint,
net forces on the magnitude of $10^{-6}$ arise.


The other type of constraint that we have imposed on this problem is
the compactness constraint that is given by enforcing zero force
outside the independently determined cell footprint $\Omega$. We see
from Fig.~\ref{COMPARE} that because of the sparse target function
$\f(x,y)$, the reconstruction of the four-pad test pattern is not
likely to be sensitive to expansions of the cell boundary.  However,
the effect of failing to enclose all of the pads within a given cell
boundary is dramatic. In Fig.~\ref{FOOTPRINT}, we performed
reconstructions where the footprint was drawn in an incorrect
orientation along the $\hat{y}$ axis, thereby failing to capture two
of the pads within the cell footprint. In these reconstructions, the
two remaining pads emerge in the reconstruction, but the effect of the
two missing pads is the spurious generation of traction forces near
the boundary of the erroneously drawn cell.

\add{Thus, the reconstruction is sensitive to both the completeness
  of the data and the footprint boundary constraint. In particular,
  for systems where the stress fields are spread about the outer
  boundary, a misspecified perimeter that misses some the interior
  domain would lead to errors, particularly if only one component of
  the displacement are given. However, if the misspecified boundary
  completely contains the cell footprint, the stress reconstruction is
  fairly robust when both components of the displacement are available.}

\subsection{Reconstruction from single cell data}

\begin{figure}[h!]
\includegraphics[width=2.8in]{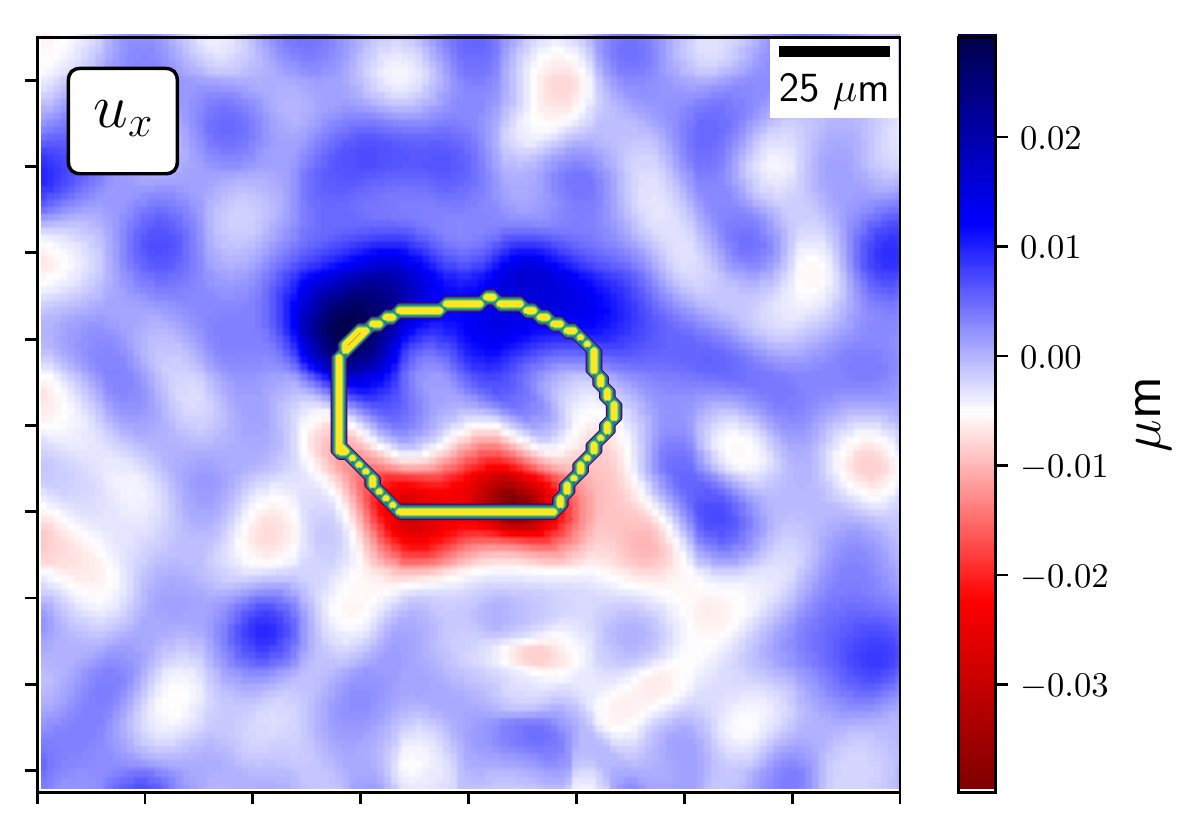}
\caption{\textbf{Mesenchymal stem cell displacement field.} Measured
  along $\hat{x}$ (the horizontal axis). Boundary of the cell (yellow)
  was hand-drawn based on bright-field image of the cell. Courtesy of
  G. Popescu (UIUC).}
\label{DATA}
\end{figure}

\begin{figure*}
\includegraphics[width=\linewidth]{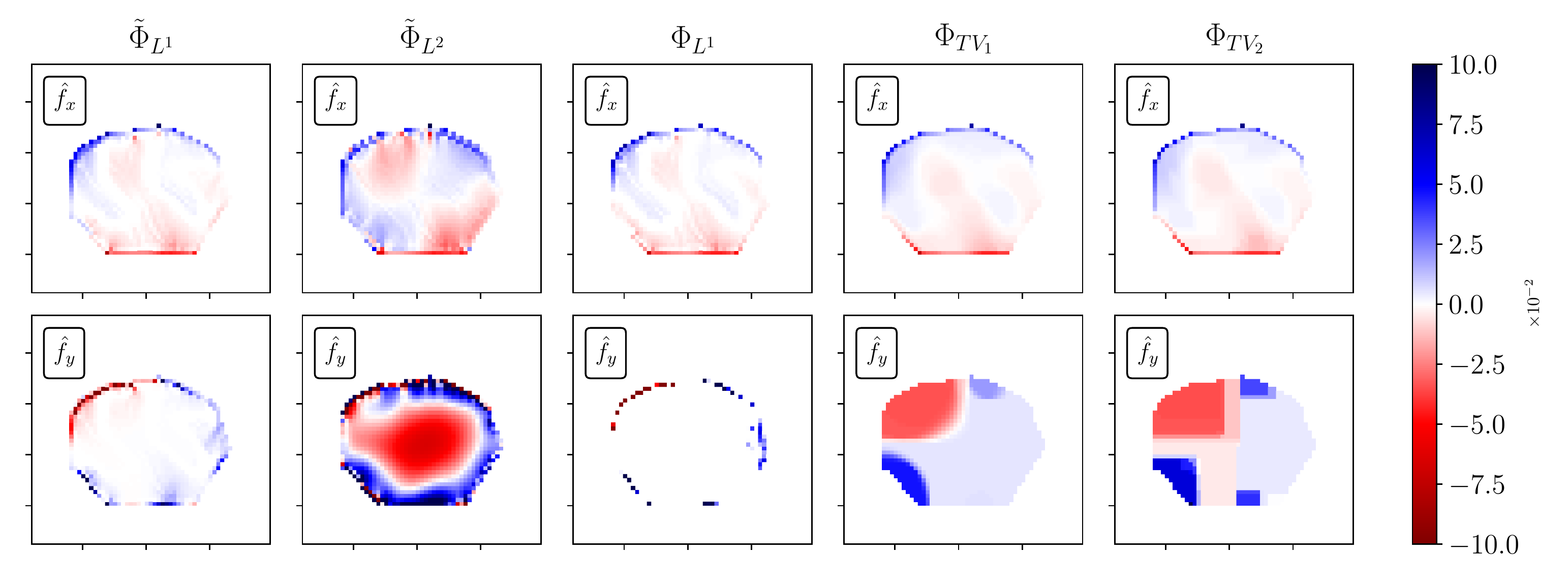}
\caption{\textbf{Reconstruction of experimental surface stress field.}
  Reconstruction of $\f$ from the measured displacements shown in
  Fig.~\ref{DATA} using the norms defined in the manuscript. In each
  case, $\lambda$ was chosen using the L-curve method as described in 
  Appendix D of the Supplementary Material.}
\label{DATA2}
\end{figure*}

To apply our method on high-resolution experimental data, we consider
the displacements resulting from stress generated by a single isolated
mesenchymal stem cell. The surface displacements were measured using
Hilbert space dynamometry which uses phase information of the periodic
signal arising from a chemically patterned grid on the substrate
\cite{POPESCU}.  In the preliminary dataset shown in Fig.~\ref{DATA},
only $x$-displacements at a resolution of the patterned grid spacing
were measured.  As we have done for the simulation data in
Fig.~\ref{COMPARE}, the $\lambda-$optimal results for the
reconstructed stress field $\hat{\f}$ using the experimental data and
the full set of constraints are shown in Fig.~\ref{DATA2}.

\begin{figure}
\includegraphics[width=2.9in]{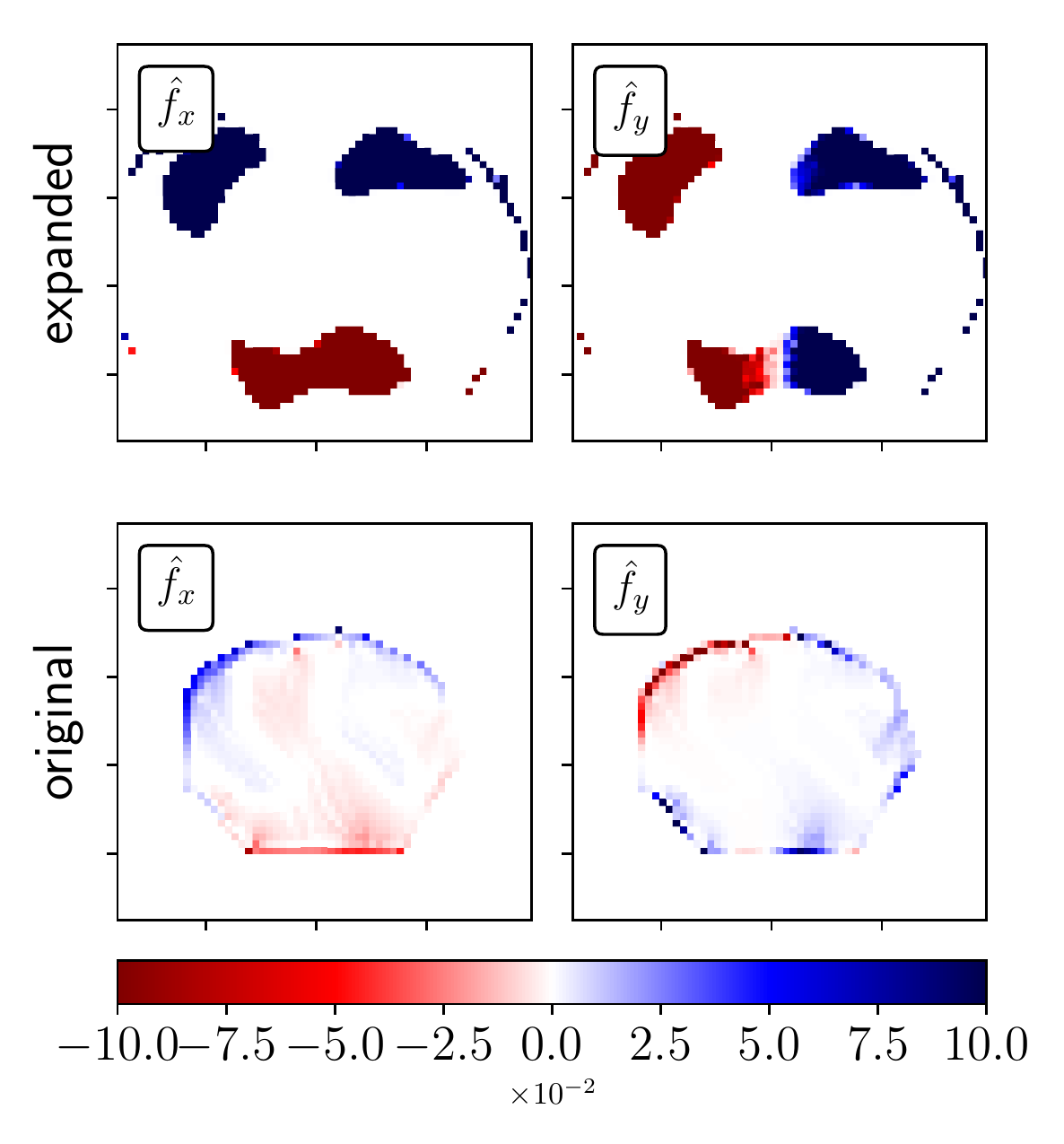}
\caption{\textbf{False footprint constraint.} Comparing the
  surface force reconstruction using the estimated cell footprint
  (from the bright field image) with the reconstruction derived from a
  false cell footprint. Since the forces are concentrated near the
  cell border, the reconstruction is sensitive to the location of the
  border (recall that only one direction of the displacement field was
  available).}
\label{DATA3}
\end{figure}

In contrast with our simulated example, these results are highly
dependent on the choice of regularization. They also illustrate the
importance of using isotropic regularization, particularly in
reconstruction problems using these type of data where only
unidirectional displacements are available.  In the reconstructions
using anisotropic penalties, the forces are qualitatively distinct
along the two directions. This behavior is undesirable as it appears
that the choice of observation axis heavily influences the outcome of
the reconstruction procedure. The isotropic $L^1$ norm, by contrast,
is robust.

As evident from the reconstructions, the surface forces are
concentrated near the border of the cell footprint. For such
boundary-dominated stress fields, the footprint constraint is expected
to be important in the recovery of $\f$. Fig.~\ref{DATA3} compares the
reconstruction with that computed with an artificially expanded
footprint. Using an incorrect footprint results in a force
distribution $\hat{\f}$ that differs from the ``true'' distribution,
especially near the borders. \added{Particularly, the qualitative
  properties of the reconstruction are markedly different when not
  imposing the true footprint.}

\added{To further probe the influence of footprint specification on
  the reconstruction of traction forces, we assumed that our
  isotropic-L$^{1}$-reconstructed ($\tilde\Phi_{L^1}$) stress field is
  the ``true'' stress field, used it to generate displacements, and
  attempted to replicate it using the expanded (and ``false'')
  cell boundary of Fig.~\ref{DATA3}. In this exercise, we added noise
  of magnitude $10^{-5} \mu$m, to approximate the noise in the real
  data.}

\begin{figure}
\includegraphics[width=2.9in]{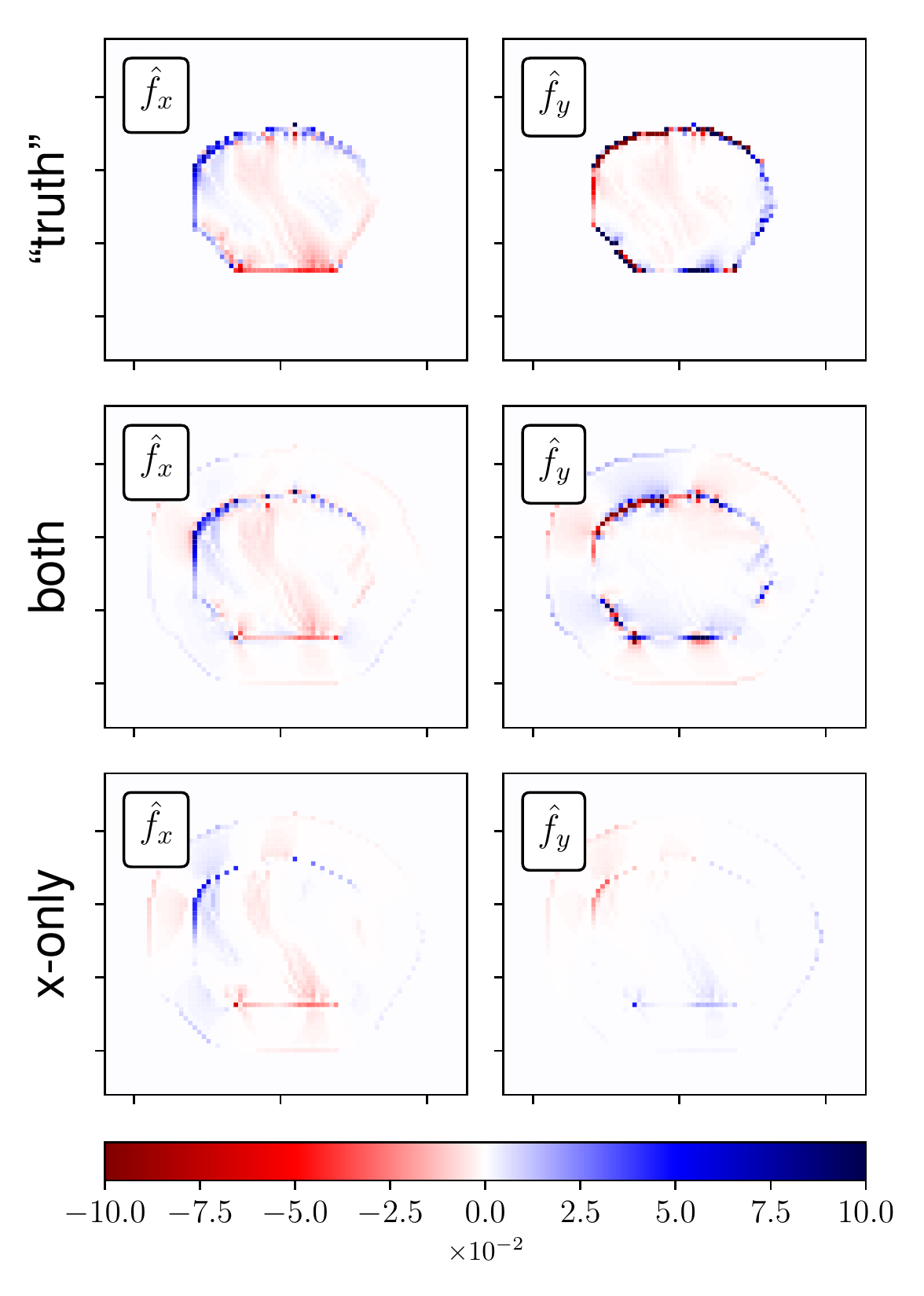}
\caption{\added{\textbf{Influence of boundary specification on force
      reconstruction.} In the top row, we take the original
    reconstruction (the first columns in Figs.~\ref{DATA2} and
    \ref{DATA3}) as the new ground ``truth.'' Using this ``true''
    stress field to generate displacements, we try to recover this
    ``true'' stress field, using a false footprint boundary. If both
    components of the displacement field are used, reasonable
    reconstruction is achieved (middle row). However, if only the
    $x$-component of the displacements are used, the reconstruction
    with the false expanded boundary constraint fails to detect the
    stress localized to the ``true'' boundary.}}
\label{DATA4}
\end{figure}

\added{The reconstruction results are shown in Fig.~\ref{DATA4}. In
  these reconstructions we assumed that we had available either both
  components of the displacement, or only a single component ($x$-only). In the
  case where both displacements are available (middle row), the
  reconstruction looks similar to that of the assumed true stress
  field, with the stress highly concentrated on the actual cell
  boundary, without explicit specification of this boundary. However,
  there is some leakage of forces outside of the actual boundary as
  well, with some forces concentrated out on the rim of the expanded
  false boundary.}

\added{When only a single component of the displacements is used, the
  reconstructed stress distribution does not gather near the actual
  cell boundary (Fig.~\ref{DATA4}, bottom row). The forces outside of the
  actual cell boundary are of similar magnitude to the forces within
  the cell in these reconstructions, particularly in the
  reconstruction of the forces in the $\hat{y}$ component. In that
  component, very little of the actual cell boundary is
  reconstructed.}

\section{Discussion}

We presented a systematic real-space approach to solving the inverse
problem associated with the reconstruction--from displacements of the
underlying substrate--of surface stresses imparted by isolated cells.
Our approach combines sparsity-favoring regularization, all
appropriate physical constraints, and an accurate piecewise affine
approximation of the exact solution to the forward problem as a system
of linear equations. This approximation to the forward problem is used
in a data mismatch term $\Phi_{\rm data}$ (Eq.~\ref{PHIDATA}).  In the
numerical implementation of the optimization problem, we also
motivated the use of a cut-off in the solution of the forward problem
that greatly reduces the rank of the inverse problem thereby
decreasing both the computational complexity of the problem and the
memory requirements. This cut-off approach is appropriate only in
scenarios in which the stress-generating cell is localized and far
from the system boundaries. Assays in which cells or layers that
extend to the boundary of the sample, or in which the substrate is of
finite depth will require\remove{d} the careful implementation of boundary
conditions defined by the sample size.

Upon further consideration of physical and geometric aspects of the
problem, we motivated additional important terms the objective
function.  The fundamental optimization problem involves minimizing an
objective function containing $L^{1}$ regularization terms that are
invariant to coordinate rotation. The anisotropy of $\hat{\f}$ derives
solely from the anisotropy in the data. Although $L^{1}$
regularization has been used in traction force microscopy
\cite{DANUSER}, in this work we also imposed a number of important
physical and geometric constraints including vanishing net force/net
torque and zero stress outside the cell footprint. Through exploration
of the mathematical features of the stress inference problem, we find
that properly identifying and implementing physical constraints (such
as no-force and no-torque) are crucial to accurate stress recovery.

We also showed the footprint boundary constraint \remove{is} can
critically impact the reconstruction, especially when adhesion sites
are distributed near the cell boundary. Such
\added{boundary-localized} focal adhesion configurations are commonly
observed in cells grown on 2D substrates. In general,
cell boundaries that artificially extend beyond the true footprint
worsens the inversion, allowing for ``leakage'' of stress beyond its
actual support. \added{These effects are especially pronounced
in reconstructions using only single-component displacement data.}

For surface stress distributions that are sparse and that arise from
localized focal adhesions, it is \added{also} important to use a
footprint that circumscribes all sources of stress. This is especially
important for cells that emanate long thin protrusions or filopodia
that may be difficult to image. Indeed, the full problem may be
extended to include the footprint as a variable in the objective
function to be minimized.  In this way, the bright field image can
also be used as data automatically infer the segmentation of the
footprint. A similar extension extension can be implemented to
reconstruct the elastic parameters if they are not known with high
certainty, resulting in a fully automated inverse problem that
utilizes both imaging and substrate displacements.



\vspace{3mm}
\noindent \textbf{Acknowledgments:} This work was supported in part
by the Intramural Research Program of the NIH Clinical Center and by grants from the
NSF (DMS-1516675) and the Army Research Office (W911NF-14-1-0472). We
are especially grateful to G. Popescu for sharing their preliminary
data on cell-induced substrate deformation measured using Hilbert
phase dynamometry.  



\section{References}

\bibliography{refs}

\onecolumngrid

\clearpage
\newpage
\setcounter{page}{1}
\begin{large}
\textbf{Supplementary Material: Mathematical Appendices}
\end{large}

\begin{appendix}

\section*{Appendix A: Elastic Green's function}

For completeness, we explicitly list the components of the Green's tensor 
for a linear elastic substrate \cite{LANDAU}

\begin{align}
G_{ss}(x,y,z) = \frac{1+\nu}{2\pi E}\left[\frac{2(1-\nu)R_{\perp}-z}{R_{\perp}(R_{\perp}-z)} + 
\frac{[2R_{\perp}(\nu R_{\perp}-z)+z^{2}]s^{2}}{R_{\perp}^{3}(R_{\perp}-z)^{2}}\right],
\tag{A1}
\end{align}
\begin{equation}
G_{zz}(x,y,z) =\frac{1+\nu}{2\pi E}\left(\frac{2(1-\nu)}
{R_{\perp}}+\frac{z^{2}}{R_{\perp}^{3}}\right),
\tag{A2}
\end{equation}
\begin{equation} 
G_{xy}(x,y,z) = G_{yx}=\frac{1+\nu}{2\pi E}\frac{[2R_{\perp}(\nu R_{\perp}-z)+z^{2}]xy}{R_{\perp}^{3}
(R_{\perp}-z)^{2}},
\tag{A3}
\end{equation}
\begin{equation}
G_{sz, zs}(x,y,z) =\frac{1+\nu}{2\pi E}\left(\frac{sz}{R_{\perp}^{3}}\pm\frac{(1-2\nu)s}{R_{\perp}
(R_{\perp}-z)}\right).
\tag{A4}
\end{equation}
where $s\equiv x,y$. The last equation with $\pm$ corresponds to
$G_{sz}$ and $G_{zs}$, respectively, and $R_{\perp} \equiv \sqrt{x^{2}
  +y^{2}}$. The Young's modulus and Poisson ratio of the elastic
substrate are denoted by $E$ and $\nu$, respectively.

\section*{Appendix B: Displacements and stresses at discrete positions}

Here, we show the explicit expressions relating displacements $\u(x_{n}, y_{m}) = \boldsymbol\Gamma \f$ at 
grid points $(x_{n},y_{m})$ in terms of stress fields at the same locations. Using 
the interpolation of $\f(x,y)$ defined by Eq.~\ref{eq:sigma_affine} in 
Eq.~\ref{eq:UMODEL1s}, we find
\begin{align}
\lefteqn{u_x(x_n,y_m) =}  \nonumber\\
 & \quad \sum_{(x_j,y_k)\in\Omega} \Bigg\{ \Bigg[f_{x}(x_j,y_k)
- x_j\left(\frac{f_{x}(x_{j+1},y_k) - f_{x}(x_{j-1},y_k)}{2\delta x}\right)
 -y_k\left( \frac{f_{x}(x_j,y_{k+1}) -
f_{x}(x_j,y_{k-1})}{2\delta y}\right) \Bigg]{\langle G_{xx} \rangle}^{nmjk} \nonumber \\
& \qquad\qquad +\left[ \frac{f_{x}(x_{j+1},y_k) -
f_{x}(x_{j-1},y_k)}{2\delta x}\right]\langle xG_{xx} \rangle^{nmjk} 
+ \left[  \frac{f_{x}(x_j,y_{k+1}) - f_{x}(x_j,y_{k-1}) }{2\delta y} \right]
\langle yG_{xx} \rangle^{nmjk}\nonumber\\
&\qquad\qquad + \Bigg[f_{y}(x_j,y_k)-x_j\left(\frac{f_{y}(x_{j+1},y_k)-
f_{y}(x_{j-1},y_k) }{2\delta x}\right)-y_k\left( \frac{f_{y}(x_j,y_{k+1}) 
-f_{y}(x_j,y_{k-1}) }{2\delta y}\right) \Bigg] \langle G_{xy} \rangle^{nmjk}\nonumber \\
&\qquad\qquad +\left[ \frac{f_{y}(x_{j+1},y_k) - f_{y}(x_{j-1},y_k) }{2\delta x}\right]  
\langle xG_{xy} \rangle^{nmjk} + 
\left[\frac{f_{y}(x_j,y_{k+1}) - 
f_{y}(x_j,y_{k-1}) }{2\delta y}\right]\langle yG_{xy} \rangle^{nmjk} \Bigg\},
\tag{B1}
\end{align}
where

\begin{align}
\langle g(x,y)G_{uv} \rangle^{nmjk} = \int_{y_k-\delta y/2}^{y_k+\delta y/2} 
\int_{x_j-\delta x/2}^{x_j+\delta x/2} g(x',y')
G_{uv}(x_n-x',y_m-y')\dd x'\dd y',\label{eq:G_ave} \tag{B2}
\end{align}
except that at the edges where we use one-sided differences so that we
are only differentiating within $\Omega$. A similar expression can be
found for solving for $u_y$ (not shown). Collecting terms, we 
write $u_{x,y}(x_{n},y_{m})$ in terms of $\f(x_{j}, y_{k})$ 
in Eq.~\ref{eq:linearsystem1}, where 

\begin{align}
\Gamma_{xx}^{nmjk} & = \langle G_{xx} \rangle^{nmjk} - 
\langle G_{xx}\rangle^{n,m,j-1,k}\frac{x_{j-1}}{2\delta x}
+ \langle G_{xx} \rangle^{n,m,j+1,k}\frac{x_{j+1}}{2\delta x} 
-  \langle G_{xx} \rangle^{n,m,j,k-1}\frac{y_{k-1}}{2\delta y}  \nonumber\\
\: &\quad + \langle G_{xx} \rangle^{n,m,j,k+1}\frac{y_{k+1}}{2\delta y} 
-\frac{\langle xG_{xx} \rangle^{n,m,j-1,k}}{2\delta x}
+\frac{\langle xG_{xx} \rangle^{n,m,j+1,k}}{2\delta x} 
- \frac{\langle yG_{xx} \rangle^{n,m,j,k-1}}{2\delta y} 
+\frac{\langle yG_{xx} \rangle^{n,m,j,k+1}}{2\delta y}, 
\label{eq:linearsystemX}\tag{B3}
\end{align}
\begin{align}
\Gamma_{xy}^{nmjk} & =  \langle G_{xy} \rangle^{nmjk} 
- \langle G_{xy} \rangle^{n,m,j-1,k}\frac{x_{j-1}}{2\delta x}
+\langle G_{xy} \rangle^{n,m,j+1,k}\frac{x_{j+1}}{2\delta x} 
- \langle G_{xy} \rangle^{n,m,j,k-1}\frac{y_{k-1}}{2\delta y} \nonumber\\
\: & \quad+\langle G_{xy} \rangle^{n,m,j,k+1}\frac{y_{k+1}}{2\delta y}  
-\frac{\langle xG_{xy} \rangle^{n,m,j-1,k}}{2\delta x}+
\frac{\langle xG_{xy} \rangle^{n,m,j+1,k}}{2\delta x} 
- \frac{\langle yG_{xy} \rangle^{n,m,j,k-1}}{2\delta y} 
+\frac{\langle yG_{xy} \rangle^{n,m,j,k+1}}{2\delta y}.
\label{eq:linearsystemY}\tag{B4}
\end{align}
Explicit closed-form expressions for the integrals in
Eq.~\ref{eq:G_ave} are given below. By defining $\Delta x_{nj}^+ = x_n -
(x_j+\delta x/2),$ $\Delta x_{nj}^- = x_n - (x_j-\delta x/2)$, $\Delta
y_{mk}^+ = y_m - (y_k+\delta y/2)$, and $\Delta y_{mk}^+ = y_m -
(y_k-\delta y/2)$, we find

\begin{align}
\langle G_{uv}\rangle^{nmjk} &=   g_{uv}( \Delta x_{nj}^+,\Delta y_{mk}^+) - 
g_{uv}( \Delta x_{nj}^+,\Delta y_{mk}^-)
-g_{uv}(\Delta x_{nj}^-, \Delta y_{mk}^+) + g_{uv}(\Delta x_{nj}^- , \Delta y_{mk}^-)\tag{B5}
\end{align}
where
\begin{align}
g_{xx}(x,y) &= \frac{\nu+1}{\pi E}\Bigg[ x(1-\nu ) \log\left( \sqrt{x^2+y^2}+y\right)
+ y \log\left(\sqrt{x^2+y^2} +x\right)-y  \Bigg] \label{eq:fxx}\tag{B6} \\
g_{yy}(x,y) &= \frac{\nu+1}{\pi E}\Bigg[ y(1-\nu ) \log\left( \sqrt{x^2+y^2}+x\right)
+ x \log\left(\sqrt{x^2+y^2} +y\right)-x  \Bigg] \label{eq:fyy}\tag{B7} \\
g_{xy}(x,y) &= -\frac{\nu(\nu+1)}{\pi E}\sqrt{x^2+y^2}. \label{eq:fxy}\tag{B8}
\end{align}
The first moments are

\begin{align}
\langle xG_{xx}(x,y) \rangle^{nmjk} & =\Big[g_{xx}(\Delta x_{nj}^+,\Delta y_{mk}^+)
- g_{xx}(\Delta x_{nj}^+,\Delta y_{mk}^-)
-g_{xx}(\Delta x_{nj}^-, \Delta y_{mk}^+)+ g_{xx}
(\Delta x_{nj}^- , \Delta y_{mk}^-)\Big] x_n\nonumber\\
\: &\qquad - \Big[g^x_{xx}( \Delta x_{nj}^+ , \Delta y_{mk}^+)- 
g^x_{xx}(\Delta x_{nj}^+ , \Delta y_{mk}^-)-
g^x_{xx}(\Delta x_{nj}^-, \Delta y_{mk}^+) + 
g^x_{xx}(\Delta x_{nj}^- , \Delta y_{mk}^-)\Big],\tag{B9}
\end{align}
where

\begin{align}
g^x_{xx}(x,y) &= \frac{\nu+1}{2\pi E} \Big[ (\nu+1)y\sqrt{x^2 + y^2}
- (\nu-1) x^2\log\left(\sqrt{x^2 + y^2} +y  \right)  \Big],\label{eq:fxxx}\tag{B10}
\end{align}

\begin{align}
\langle yG_{xx}(x,y) \rangle^{nmjk} & = 
\Big[g_{xx}(\Delta x_{nj}^{+},\Delta y_{mk}^+)-g_{xx}(\Delta x_{nj}^+,\Delta y_{mk}^-)
-g_{xx}(\Delta x_{nj}^{-}, \Delta y_{mk}^+) + 
g_{xx}(\Delta x_{nj}^{-}, \Delta y_{mk}^-)\Big] y_m\nonumber\\
\: &\qquad - \Big[ g^y_{xx}(\Delta x_{nj}^+,\Delta y_{mk}^+) 
- g^y_{xx}(\Delta x_{nj}^+,\Delta y_{mk}^-)
- g^y_{xx}(\Delta x_{nj}^-,\Delta y_{mk}^+) 
+ g^y_{xx}(\Delta x_{nj}^-,\Delta y_{mk}^-)\Big],\tag{B11}
\end{align}
where

\begin{align}
g^y_{xx}(x,y) &=\frac{\nu+1}{2\pi E} \Bigg[y^2\log\left(\sqrt{x^2+y^2}+x \right)
-\sqrt{x^2+y^2}\left((2\nu-1)x + \frac{1}{2}\sqrt{x^2+y^2} \right)  \Bigg], \label{eq:fyxx}\tag{B12}
\end{align}
and

\begin{align}
\langle xG_{xy}(x,y)\rangle^{nmjk} & = 
\Big[g_{xy}(\Delta x_{nj}^{-},\Delta y_{mk}^+)-g_{xy}(\Delta x_{nj}^{-},\Delta y_{mk}^-)
- g_{xy}(\Delta x_{nj}^{+},\Delta y_{mk}^+) + g_{xy}(\Delta x_{nj}^+,\Delta y_{mk}^-) \Big]x_n \nonumber\\
\: & \qquad -\Big[g_{xy}^x(\Delta x_{nj}^{+},\Delta y_{mk}^+) - g_{xy}^x(\Delta x_{nj}^{+},\Delta y_{mk}^-)
-g_{xy}^x(\Delta x_{nj}^{-},\Delta y_{mk}^+)+g_{xy}^x(\Delta x_{nj}^{-},\Delta y_{mk}^-)\Big]\tag{B13}
\end{align}
where

\begin{align}
g_{xy}^x(x,y) &=\frac{\nu(\nu+1)}{\pi E}\Big[ \frac{y^2}{2}\log\left(\sqrt{x^2+y^2} +x\right)
-\frac{1}{4}\sqrt{x^2+y^2}\left(\sqrt{x^2+y^2}+2x\right) \Big]. \label{eq:fxxy}\tag{B14}
\end{align}
Analogous expressions are straightforwardly derived for
$\Gamma_{yx}^{nmjk}$ and $\Gamma_{yy}^{nmjk}$.

All of the above expressions may be found through direct iterated
evaluation of the integrals, as long as $n\neq m$ or $j\neq k$ the
integrand (effectively the Green's function) is bounded, hence making
Fubini's theorem applicable given the compactly supported domains of
integration.

In the special case where $n=m$ and $j=k$, these formulae also
hold. This fact is found by decomposing the integration domain to
exclude the origin, for instance in the manner
\begin{equation}
\int_{ -\Delta y/2 }^{\Delta y/2 }\int_{- \Delta x/2 }^{\Delta x/2 } \d\r = 
\lim_{\varepsilon\to0} \left( \int_\varepsilon^{\Delta y/2}  
+  \int_{-\Delta y/2}^\varepsilon   \right)\int_{\Delta x/2}^{\Delta x/2} \d\r.~\label{eq:splitdomain}\tag{B15}
\end{equation}
Since the antiderivatives of
Eqs.~\ref{eq:fxx},~\ref{eq:fyy},~\ref{eq:fxy},~\ref{eq:fxxx},~\ref{eq:fyxx},
and~\ref{eq:fxxy} all have well-defined limits with only removable
discontinuities at the origin, integrals of the Green's functions
defined through Eq.~\ref{eq:splitdomain} all converge about the origin
and the equations above also hold in the case where $n=m$ and $j=k$.
These explicit expressions allow us to accurately evaluate
$\u(\r_{i})$ in $\Phi_{\rm data}[\f]$.

\section*{Appendix C: Decay of displacement fields}

Note that $u_x$ and $u_y$ are symmetric in form. Hence, it will
suffice to prove just one of these assertions. Eq.~\ref{eq:UMODEL1s}
can be written as

\begin{align}
u_x(\r) & = \frac{1+\nu}{\pi E} \int \frac{\dd \r}{|\r-\r'|} 
\Bigg\{ \left[ \frac{\nu(x - x')^2}{|\r-\r'|^2} + 1-\nu \right]f_{x}(\r') +
\nu\frac{(x-x')(y-y')}{|\r-\r'|^2} f_y(\r')  \Bigg\} \nonumber\\
& \equiv \frac{1+\nu}{\pi E}   \int\frac{\rho_x(\r,\r')}{|\r-\r'|} \dd \r' \label{eq:rhoeq}\tag{C1}
\end{align}
where $\rho_x(\r,\r')$ is $\mathcal{O}(1)$ as
$\vert\r\vert\to\infty$. Without loss of generality, we assume that
the coordinate system is centered at some point
$\mathbf{0}\in\Omega$. The Euclidean distances can then be represented
through the binomial expansion,

\begin{align}
\frac{1}{|\r-\r'|^p} &= \frac{1}{|\r|^p}\frac{1}{\left(1 - \frac{2\r\cdot\r'}{|\r|^2} 
+\frac{\vert\r'\vert^{2}}{\vert\r\vert^{2}}\right)^{p/2}} = 
\frac{1}{|\r|^p}\sum_{k=0}^\infty {{{p\over 2}+k-1}\choose{k}} 
{\left( \underbrace{ \frac{2\r\cdot\r' - |\r'|^2}{|\r|^2}}_{\mathcal{O}(|\r|^{-1})} \right)}^k.
\tag{C2}
\end{align}
Since $\r\not\in\Omega$ and $\r'\in\Omega$, the series converges in
the $\vert \r\vert \to \infty$ limit. Plugging this series into the
last line of Eq.~\ref{eq:rhoeq}, where $p=1$, one sees that it in order to show
that the magnitude of $u_x(\r)$ is $\mathcal{O}(|\r|^{-q})$, it
suffices to show that $\int \rho(\r,\r')\dd\r' \leq
{\cal O}(|\r|^{-q+1})$.

Using the fact that $\int \f(\r)\d\r = \mathbf{0}$, one finds that
\begin{align}
\int \rho_{x}(\r,\r')\d\r'  & =  \int (1-\nu)f_x(\r')\d\r'
+\nu\int\left[\frac{(x - x')^2}{|\r-\r'|^2} f_{x}(\r') +
\frac{(x-x')(y-y')}{|\r-\r'|^2} f_y(\r')  \right]\d\r' \nonumber\\
\: &= \frac{\nu}{|\r|^2}\int\Bigg[(x - x')^2f_{x}(\r')
+ (x-x')(y-y')f_y(\r')  \Bigg]
\sum_{k=0}^\infty \left[\frac{2\r\cdot\r' - |\r'|^2}{|\r|^2} \right]^k \d\r'.
\tag{C3}
\end{align}
Expanding the leading order term of this expression, we see that

\begin{align*}
\int\rho_{x}(\r,\r')\d\r 
&= \frac{\nu}{|\r|^2}\int\Bigg[(x- x')^2f_{x}(\r')+(x-x')(y-y')f_y(\r')\Bigg]
\d\r' \nonumber\\
\: &= \frac{\nu}{|\r|^2}  \Bigg[  -2x\int x' f_x(\r')\d\r'  + \int x'^2 f_x(\r')\d\r'- x\int y' f_y(\r') \d\r'
- y\int x' f_y(\r')\d\r'
+ \int x'y' f_y(\r') \d\r'\Bigg] \nonumber \\
\: & =\mathcal{O}(|\r|^{-1}).\tag{C4}
\end{align*}
Hence, it is evident that this integral \added{is} of $\mathcal{O}(|\r|^{-2})$,
where to the leading order we have

\begin{align}
\lefteqn{u_x(\r) = \frac{1+\nu}{\pi E |\r|^2}
\Bigg[ -2\nu\frac{x}{|\r|}\int x' f_x(\r')\d\r'
-\frac{x\nu}{|\r|}\int y' f_y(\r') \d\r'  }\nonumber\\
&\qquad- \frac{y\nu}{|\r|}\int x' f_y(\r')\d\r' 
+ (1-\nu)\frac{\r}{|\r|}\cdot\int \r' f_x(\r') \d\r'\Bigg] + \mathcal{O}(|\r|^{-3}).
\tag{C5}
\end{align}

\section*{Appendix D: Choice of regularization penalty parameter $\lambda$}

The ``L-curve'' below illustrates the optimal choice of $\lambda$ for each reconstruction
used in this paper. 

\begin{suppfigure}[h]
\includegraphics[width=\linewidth]{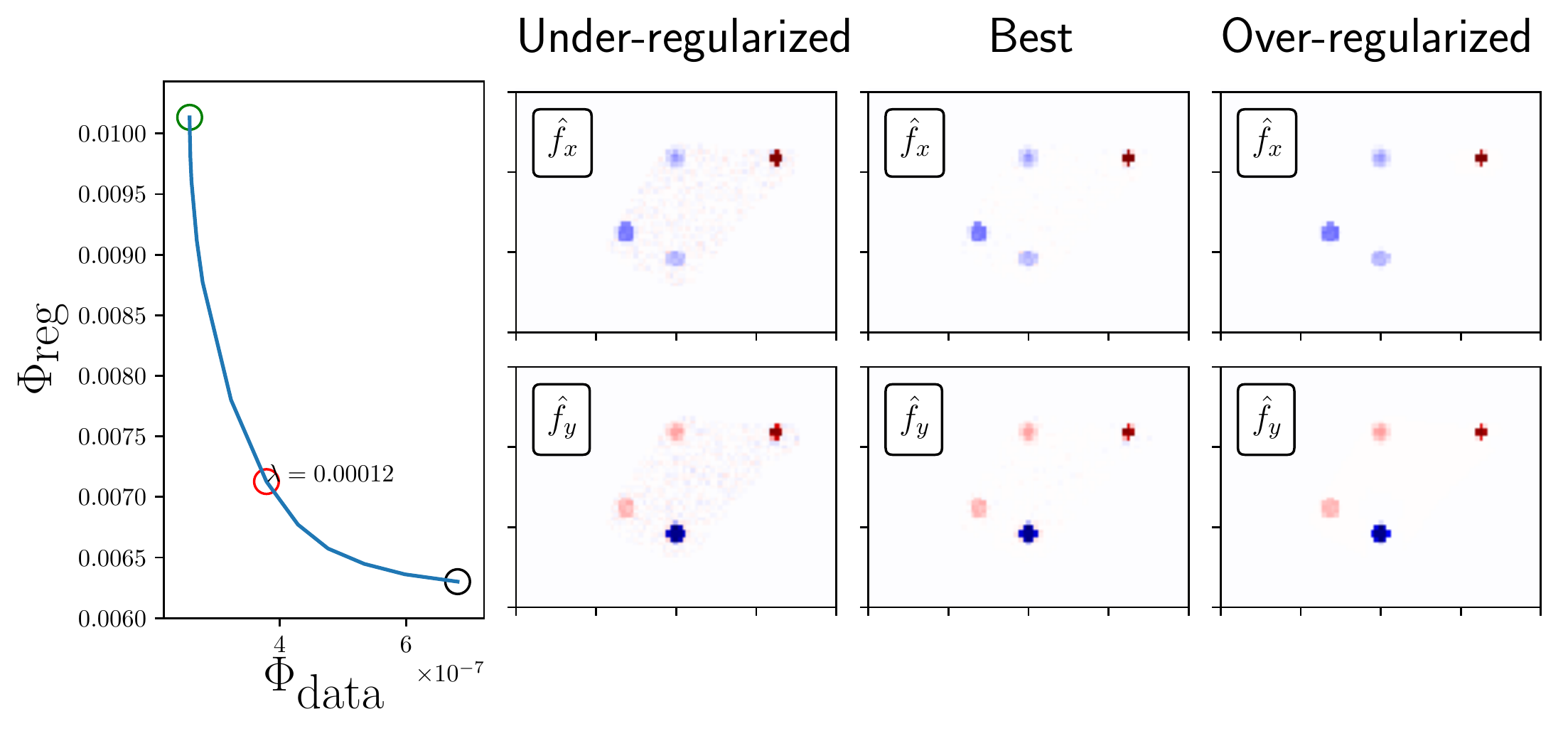}
\caption{\textbf{Choice of $\lambda$} by plotting the tradeoff between
  regularity and data mismatch for different values of $\lambda$. The
  ``optimal'' value (red circle) is chosen by to be the point farthest
  away from the line segment joining the two ends of the plot (green
  and black circles).  Reconstructions under the different values of
  $\lambda$ given by these circles are shown. The green circle
  corresponds to solution with low regularity and is hence
  ``under-regularized.'' The black circle coincides with a solution of
  high regularity and is hence ``over-regularized.''}
\label{LAMBDA}
\end{suppfigure}

\begin{suppfigure}[h]
\includegraphics[width=0.95\linewidth]{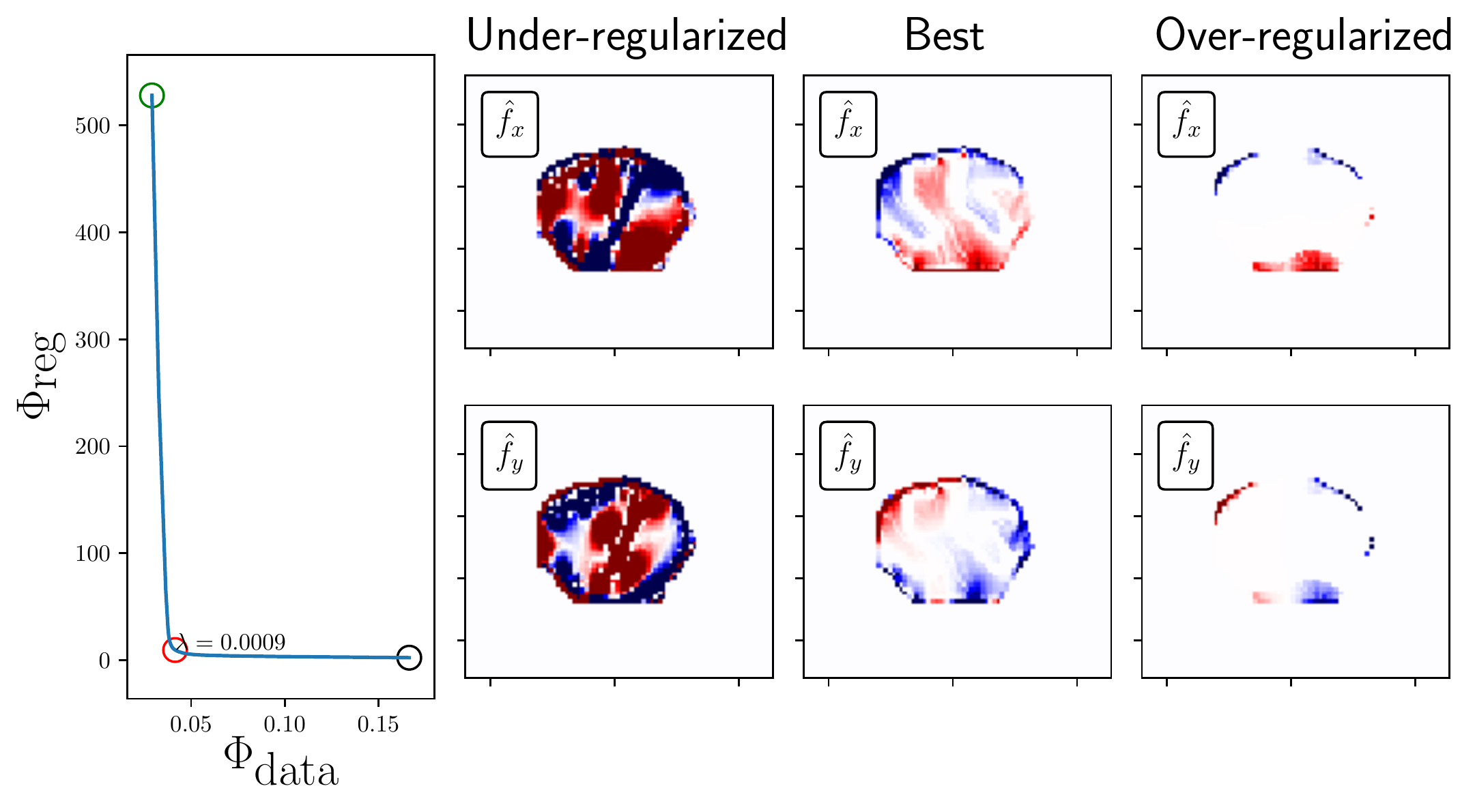}
\caption{\added{\textbf{Choice of $\lambda$} for mesenchymal cell data
    (where displacements are known only in one direction). The optimal
    reconstruction corresponds to the elbow of the tradeoff
    curve. Color contrast enhanced relative to scale used in the
    manuscript body.}}
\label{LAMBDA2}
\end{suppfigure}

\end{appendix}
\end{document}